\documentclass{article}
\input{tcilatex}

\begin{document}

\section{\textbf{Isotope dependence of band-gap energy.}}

\subsubsection{\textbf{V.G. Plekhanov}$^{\bullet }$\textbf{\ and N.V.
Plekhanov}}

\textbf{Computer Science College, Erika Street 7a,Tallinn 10416, ESTONIA}

\textbf{Abstract}. The results of the quantitative investigations of the
renormalization of the absorption edge of different compounds by the isotope
effect are described. The obtained dependence of the band gap energy versus
isotope mass is in reasonable agreement with estimated value obtained from
the temperature dependence of E$_{g}$ in LiH crystals. For the first time it
was shown that the dependence E$_{g}$ = f(M) for the different materials has
a parabolic character ln($\partial $E$_{g}/\partial M$) = 6.105(lnE$_{g}$)$%
^{2}$ - 7.870(lnE$_{g}$) + 0.565. Nonlinear character of this dependence may
be indicated on the nonlinear dependence of the potential scattering on the
isotope mass.

\textbf{PACS: 32.10.B + 71.20.P + 78.20 + 78.40.F}

Isotope substitution is a well-defined and easily controllable method to
investigate intrinsic renormalization mechanisms of elementary excitations
of solids. First of all, phonon frequencies are directly affected by changes
of the average mass of the whole crystal or its sublattices (in the case of
virtual crystal approximation) even when we look upon them as noninteracting
particles, i.e. as harmonic oscillators. The renormalization of the
fundamental electronic gap by electron-phonon interaction also depends on
the isotope mass. Measuring the energy gaps in samples with different
isotopic composition then yields the difference in the changes of the
valence- and conduction-band renormalization.

In this communication we report the first results of the quantitative study
the dependence of the band-gap energy on the isotope effect for different
compounds. As was mentioned above isotopic substitution only affects the
wavefunctions of phonons; therefore, the energy values of electron levels in
the Schrodinger equation ought to have remained the same. This, however, is
not so, since isotopic substitution modifies not only the phonon spectrum,
but also the constant of electron-phonon interaction. It is for this reason
the energy values of purely electron transitions in molecules of hydride and
deuteride are found to be different (see, e.g. [1]). This effect is even
more prominent when we are dealing with a solid [2]. This conclusion was
confirmed on a qualitative level as early as the 1930s in Ref. [3].
Intercomparison of absorption spectra for thin films of LiH and LiD at room
temperature revealed [3] that the longwave maximum (as we now know to be the
exciton peak (see, e.g. [2])) moves 64.5 meV towards the shorter wavelengths
when H is replaced by D. As will be shown below, this effect becomes even
more pronounced at low temperatures.

$^{\bullet }$\textbf{Correspondence to V.G. Plekhanov; \TEXTsymbol{<}e-mail%
\TEXTsymbol{>} vgplekhanov@iati.ee}

The mirror reflection spectra of mixed and pure LiD crystals cleaved in
liquid helium are presented in figure 1. For comparison, on the same diagram
we have also plotted the reflection spectra of LiH crystals with clean
surfaces. All spectra have been measured with the same apparatus under the
same conditions. As the deuterium concentration increases, the long-wave
maximum broadens and shifts towards shorter wavelengths. As can clearly be
seen in figure 1, all spectra exhibit a similar long-wave structure. This
allows us to attribute this structure to the excitation of the ground (1s)
and the first excited (2s) exciton states. The energy values of exciton
maxima for pure and mixed crystals at 2K are presented in Table I. The
binding energies of excitons E$_{b}$, calculated by the hydrogen-like
formula, and the energies of interband transitions E$_{g}$ (details see [4])
are also given in Table I.

The ionization energy, found from the temperature quenching of the peak of
reflection spectrum [4] of the 2s state in LiD is 12 meV. This value agrees
fairly well with the value of $\Delta $E$_{2s}$ calculated by the
hydrogen-like formula. Moreover, E$_{b}$ = 52 meV for LiD agrees well with
the energy of activation for thermal quenching of free-exciton luminescence
in these crystals [2].

At the weak scattering potential the mean-square vibrational amplitude 
\TEXTsymbol{<}u$^{2}$\TEXTsymbol{>} of an atom dependes on the phonon
frequencies and the eigenvectors, the atomic masses (at low temperatures),
as well as the temperature (at high temperature). Isotope substitution
results in a slightly different vibrational amplitudes (epecially at low
temperature) and phonon frequencies and in first approximation equivalent to
changing the temperature (see also [5]). The mass dependence of \TEXTsymbol{<%
}u$^{2}$\TEXTsymbol{>} becomes vanishingly small at temperature on the order
or higher than the Debye temperature. Changes in either the isotope masses
or temperature thus lead to changes in the band gap (see above Fig. 1) via
the electron-phonon interaction, even at zero temperature in the case of
mass changes. In general, the renormalization of the band gap [4] and its
temperature dependence result from a complicated interplay of 1) first- and
second-order electron-phonon interactions that contribute to the energy of
conduction and valence bands; 2) changes due to thermal or isotopic lattice
expansion, and 3) changes in the phonon occupation numbers (see [5] and
references therein).

The dependence of the band gap energy on isotopic composition has already
been observed for insulators and lowest (indirect-direct) gap of different
semiconductors (see also [6]). It has been shown to result primarily from
the effect of the average isotopic mass on the electron-phonon interaction,
with a smaller contribution from the change in lattice constant. This
simplest approximation, in which crystals of mixed isotopic composition are
treated as crystals of identical atoms having the average isotopic mass is
reffered to as virtual crystal approximation (VCA) [7]. Going beyond the
VCA, in isotopically mixed crystals one would also expect local fluctuations
in the band gap energy from statistical fluctuations in local isotopic
composition within some effective volume, such as that of an exciton. As
follows from Fig. 1 excitons in LiH$_{x}$D$_{1-x}$ crystals display a
unimodal character, which facilitates the interpretation of their
concentration dependence. Fig. 2 shows the concentration dependence of the
energy of interband transition E$_{g}$. As can be seen from Fig. 2 VCA
method cannot describe observed experimental results. As will shown below
this deviation from linear low (VCA approximation) is connected with
isotope-induced-disorder in isotope mixed crystals LiH$_{x}$D$_{1-x}.$

The temperature and isotopic mas dependence of a given energy gap E$_g$(T,M$%
_i$) can be described by average Bose-Einsteinm statistical factor n$_B$
corresponding to an average phonon frequency $\theta _i$ as (see also [5; 8]

E$_g$(T,M$_i$) = E$_{bar}$ - a$_r$($\frac{M_{nat}}{M_i}$)$^{1/2}\left[ \text{%
1 + 2n}_{\text{B}}\right] $, (1)

where n$_{B}$ \ = 1/[exp$\left( \frac{\theta _{i}}{\text{T}}\right) $ -1]
and E$_{bar}$ and a$_{r}$ the unrenormalized (bare) gap and the
renormalization parameter, respectively. In the low-temperature limit, T%
\TEXTsymbol{<}\TEXTsymbol{<}$\theta _{i}$, equation (1) reduces

E$_g$(T,M$_i$) = E$_{bar}$ - a$_r$($\frac{M_{nat}}{M_i}$)$^{1/2}$ (2)

Here E$_{g}$(T,M$_{i}$) is independent of temperature and proportional to
(1/M$_{i}$)$^{1/2}$, whereas a$_{r}$ $is$ the energy difference between the
unrenormalized gap (M$_{i}$ $\rightarrow \infty $) and the renormalized
value [5].

In the high-temperature limit, T\TEXTsymbol{>}\TEXTsymbol{>}$\theta _i$ and
Eq. (1) can be written as

E$_g$(T,M$_i$) = E$_{bar}$ - 2T$\frac{a_r}\theta $, (3)

and E$_g$(T,M$_i$) is independent of M$_i$ [5]. The extrapolation of Eq. (3)
to T = 0K can be used to determine the unrenormalized gap energy E$_{bar}$ ,
i.e., the value that corresponds to atoms in fixed lattice position without
vibrations (frozen lattice [4]), from the measured temperature dependence of
E$_g$(T) in the high-temperature (i.e. linear in T) region.

Using Eq. (2) it can be written the difference in energy $\Delta $E$_g$
between a given energy gap in isotopically pure material (LiH) and its
isotope analogue (LiD)

$\Delta $E$_g$ = E$_g$(M$_i$) - E$_g$(M$_{nat}$) = a$_r$ $\left[ \text{1 - }%
\left( \frac{M_{nat}}{M_i}\right) ^{1/2}\right] $, (4)

As can be seen from Table 1 and results of [5] $\Delta $E$_{g}$ at 2K equals 
$\Delta $E$_{g}$ = 0.103 eV and E$_{g}$(LiH, T = 0K) =5.004 eV (linear
approximation and E$_{g}$(LiH, T = 300K) = 4.905 eV then using Eq. (3) we
obtain a$_{r}$ = 0.196 eV. This magnitude is very close (approximately 84\%)
to the value of 0.235 eV of zero vibration renormalization of the energy
band gap in LiH crystals (details see [4;5]). Using Eq. (4) we obtain $%
\Delta $E$_{g}$(theor) = 0.134 eV that is very close, on the other hand, to
observed experimental value equals 0.103 eV. The discrepancy between these
values may be caused by the negligible contribution of the isotopic lattice
expansion to the band gap renormalization.

The single-mode nature of the exciton reflection spectra of mixed crystals
LiH$_{x}$D$_{1-x}$ agrees qualitatively with the results obtained with the
virtual crystal model (see, e.g. [7; 9]), being at the same time its extreme
realization, since the difference between ionization potentials ($\zeta $)
for this compounds is zero. According to the virtual crystal model, $\zeta $
= 0 implies that $\Delta $E$_{g}$ = 0, which in contradiction with the
experimental results for LiH$_{x}$D$_{1-x}$ crystals (details see, also
[6]). In the light of obtained dependence E$_{g}$ = f(T, M$_{i}$) for LiH$%
_{x}$D$_{1-x}$ it is very interested to study such dependence for different
materials. Although the details of the renormalization process of E$_{g}$
are not known, observation of the isotope renormalization of E$_{g}$ for
many materials force to pay a great attention on this effect.

By now the change in E$_{g}$ caused by isotopic substitution has been
observed for many broad-gap and narrow-gap compounds. Below we briefly
discuss the variation of the electronic gap (E$_{g}$) of different crystals
with its isotopic composition. In the last decade the whole row of different
semiconducting crystals was grown. These crystals are diamond [10], copper
halides [11; 12], germanium [13;14], Si [8] and GaAs [15]. All enumerated
crystals show the dependence of the electronic gap on the isotope masses. It
should be noted that the indicated effect (the variation of E$_{g}$ and E$%
_{b}$, see Table I and II) have maximum values in LiH crystals, although
this effect in other crystals with isotopic composition are currently being
reliably measured and investigated well. All numerated crystals show the
dependence of the electronic gap on the isotope masses.

Before we complete the analysis of these results we should mention that
before these investigations, studies were carried out on the isotopic effect
on exciton states for a whole range of crystals by Kreingol'$d$ and
coworkers [16-21]. First of all we should name the classic crystals Cu$_{2}$%
O [17-19] with the substitution O$^{16}$ $\rightarrow $ O$^{18}$and Cu$^{63}$
$\rightarrow $ Cu$^{65}$. Moreover, there have been some detailed
investigations of the isotopic effect on ZnO crystals [6] where E$_{g}$ was
seen to increase by 55 cm$^{-1}$ (O$^{16}\rightarrow $ O$^{18}$) and 12 cm$%
^{-1}$ ( at Zn$^{64}$ $\rightarrow $ Zn$^{68}$) [16;20]. In the paper of
Kreingol'd and coworkers [21] it was shown that the substitution of a heavy S%
$^{34}$ isotope for a light S$^{32}$ isotope in CdS crystals resulted in a
decrease in the exciton Rydberg constant (E$_{b}$ ).

More detailed investigations of the exciton reflectance spectrum in CdS
crystals were done by Zhang et al [22]. Zhang et al to directly determine
the binding energy (E$_b$) and the corresponding band gaps ($E_g$) from a
hydrogenlike model.

In this manner it was obtain binding energies of A excitons of 26.4$\pm $%
0.02 meV and 26.8$\pm $0.02 meV in $^{112}$CdS and $^{nat}$CdS,
respectively. The corresponding band gaps E$_g$($\Gamma _{7c}$ - $\Gamma
_{9v}$) are 2.5806(2) eV and 2.5809(2) eV, respectively. In the case of B
excitons, these values are E$_b$ = 27.1$\pm $0.2 meV (27.1$\pm $0.2 meV) and
E$_g$ ($\Gamma _{7c}$ - $\Gamma _{7v}$) = 2.5964(2) $\left[ 2.5963(2)\text{eV%
}\right] $ for the $^{112}$ CdS $\left[ ^{nat}\text{CdS}\right] $ sample.
Unfortuntealy, the n =2 excited states of the A and B excitons could not be
observed in other isotopic CdS samples. Better samples are required for such
measurements.

For GaAs or ZnSe, isotope substituents of either type should lead to shifts
of the E$_{0}$ gap which have been calculated to be 430 (420) and 310 (300) $%
\mu $eV/amu for cation (anion) mass replacement, respectively [15;23]. These
values are in reasonable agreement with data measured for GaAs $\left[
\partial E_{0}\text{/}\partial \text{M}_{Ga}\text{ = 390 (60) }\mu eV/amu%
\right] $ and preliminary results for isotopic ZnSe obtained by Zhang et al
[22] based on photoluminescence measurements of the bound exciton (neutral
acceptor I$_{1}$) $\left[ \partial \text{E/}\partial \text{M}_{Se}\text{ =
140}\pm \text{ 40 }\mu \text{eV/amu and }\partial \text{E/}\partial M_{Zn}%
\text{ = 240}\pm \text{ 40 }\mu eV\text{/amu}\right] .$

Such behavior, however, is not found in wurtzite CdS. A previous
reflectivity and photoluminescence study of $^{nat}$Cd$^{32}$S and $^{nat}$Cd%
$^{34}$S shows [21] that for anion isotope substitution the ground state (n
= 1) energies of both A and B excitons have a positive energy shift with the
large rate of $\partial $E/$\partial M_S$ = 740$\pm $100 $\mu $eV/amu. This
value is more than one order of magnitude larger than $\partial $E/$\partial 
$M$_{Cd}$ obtained by Zhang et al [22].

Several groups have conducted low-temperature studies of the direct and
indirect band gaps of natural and isotopically controlled Ge single
crystals. For the first time Agekyan et al [13] used photoluminescence,
infrared absorption, and Raman spectroscopy with a Ge crystals of natural
composition and a crystals with 85 \% $^{76}$ Ge and 15\% $^{74}$Ge. They
found an indirect band-gap change $\Delta $E$_{g}$ = 0.9 meV and a direct
band-gap change $\Delta $E$_{g}$ = 1.25 meV with an error of $\pm $0.05 meV.
Etchegoin et al [24] and Davies et al [25] reported photoluminescence
studies of natural and several highly enriched , high quality single
crystals of Ge. Measurement of the energies of impurity-bound excitons
permits by Davies et al the direct determination of band-gap shifts with the
crystal isotope mass because the radiative recombination does not require
phonon participation. As may be expected from the very large Bohr orbit of
the excitons (see Davies et al [25] and reference therein), their binding
energy only depends on the average isotope mass and not on the isotopic
disorder (see, however [4]). The rate of band-gap energy change with isotope
mass as determined by Davies et al is dE$_{IG}$/dA = dE$_{NP}$/dA = 0.35$\pm 
$0.02 meV/amu. Etchegoin et al [24] obtained a very similar value.

Zollner et al.[25] have performed a numerical calculation of the electronic
bands using an empirical pseudopotential method including the necessary
lattice dynamics. They found for Ge (dE$_{IG}$/dA)$_{e-p}$ = 0.41 meV. The
total calculated shift of the indirect band-gap energy with isotope mass
adds up to (dE$_{IG}$/dA)$_{total}$ = 0.48 meV. This result compares
favorably with the experimental values stated above by Davies et al and by
Etchegoin et al who reported (dE$_{IG}$/dA)$_{total}$ = 0.37$\pm $0.01
meV/amu.

Measurements of the direct band gap at the $\Gamma $ point ($\vec{k}$ = 0)
in the Brillouin zone have also been performed. Though the direct band gap
is technologically less important than the minimum indirect band gap,
determing the dependence of this gap on isotope mass is of the same
fundamental significance as the indirect band-gap studies. Davies et al [27]
used low-temperature optical-absorption measurements of very thin samples of
Ge single crystals with natural composition and three different, highly
enriched isotopes. They found dE/dA = 0.49 $\pm $ 0.03 meV/amu for the
temperature extrapolated to zero. Parks et al [28] have used piezo- and
photomodulated reflectivity spectra of four monoisotopic and one natural Ge
crystals. These techniques do not require the extreme sample thinning which
is necessary for optical-absorption measurements and the derivative nature
of the spectra emphasizes the small changes. The excellent signal-to- noise
ratio and the superb spectral resolution allowed a very accurate
determination of the dependence of E$_{DG}$ on isotopic mass. At very low
temperatures an inverse square-root dependence accurately describes band-gap
dependence \ E$_{DG}$ = E$_{DG}^{\infty }$ + $\frac{C}{\sqrt{M}}$ .

A fit through five data points yields E$_{DG}^{\infty }$ = 959 meV and C =
-606 meV/amu$^{1/2}$. Written as a linear dependence for the small range of
isotopic masses, Parks et al find dE$_{DG}$/dA = 0.49 meV/amu, in perfect
agreement with the results of Davies et al [27]. Parks et al also determined
the isotope mass dependence of the sum of the direct gap and the split-off
valence band $\left( \Delta _{0}\right) $ and find d(E$_{DG}$ + $\Delta _{0}$%
)/dA = 0.74 meV/amu. The experimental results can be compared to the Zollner
et al [26] calculations which are found to be of the correct order of
magnitude. The theoretical estimates for the contributions of the linear
isotope shifts of the minimum, indirect gaps which are caused by
electron-phonon interaction, are too large by a factor of $\sim 1.7$ and for
the smallest direct gap they are too large by a factor $\sim $ 3.2.

Substitution of Ga$^{70}$ on the Ga$^{76}$ increases the band gap in GaAs
[23] on 10.5 cm$^{-1}$. The interested results were communicated in papers
of Cardona and coworkers [15; 12], where it was studied the dependence of E$%
_{g}$ on the isotope effect in CuCl crystals. When the Cu$^{64}$ on the Cu$%
^{65}$ is substituted the value of E$_{g}$ in CuCl crystals decreased on
1.24 cm$^{-1}$, e.g. the isotope effect on the electronic excitation has an
opposite sign.

Considering the series of Ge, GaAs, ZnSe, CuBr, for example, the 3d states
of the first constituent play an increasing role in determining the band
structure. In Ge these states can be considered as localized core states
(atomic energy level $\approx $ -30 eV). Already, however, in GaAs they have
moved up in energy by 10 eV, and their hybridization with the top of the
valence band affects the gap . Proceeding further in the series, this effect
becomes more important, and in CuBr and Cu 3d states even overlap in energy
with halogen p-states, with which they strongly hybridize. Therefore, we
cannot exclude that the main reason for the opposite sign of the isotopic
effect in these compounds may be connected to the different character of the
d-electron-phonon interactions in these semiconductors [6].

Analogous investigations of the direct absorption edge of Si (E$_{g}$ =
3.562 eV) have been performed in paper [8] made the value of the coefficient 
$\partial $E$_{g}$/$\partial $M = 2.0 meV/amu. The measurements of $\partial 
$E$_{g}$/$\partial $M for isotope effect in CsH (CsH$\rightarrow $CsD) gave
the value $\partial $E$_{g}$/$\partial $M = 60 meV/amu [29]. The change of
the indirect gap of diamond between pure C$^{12}$ and C$^{13}$has been
determined by Collins et al [10], using for this purpose the
cathodoluminescence spectra of diamond. From the results of Collins et al it
was concluded that the dominant contribution arises from electron-phonon
coupling, and that there is a smaller contribution due to a change in volume
of the unit cell produced by changing the isotope. These two terms were
calculated as 13.5 $\pm $ 2.0 and 3.0 $\pm $ 1.3 meV respectively (details
see, also [30]).

All of these results are documented in Table II, where the variation of E$_g$
and $\partial $E$_g$/$\partial $M are shown at the isotope effect. We should
highlighted here that the most prominent isotope effect is observed in LiH
crystals, where the dependence of E$_b$ = f (C$_H$) is also observed and
investigated (see, also [2]). Using the least-squares method it was found
the empirical dependence of ln$\partial $E$_g$/$\partial $M $\sim $ f(lnE$_g$%
), which is depicted on Fig. 3. As can be seen the indicated dependence has
a parabolic character:

ln($\partial $E$_g/\partial M$) = 6.105(lnE$_g$)$^2$ - 7.870(lnE$_g$) +
0.565. (5)

From this figure it can be concluded also that the small variation of the
nuclear mass causes the small changes in E$_{g}$ also. When the nuclear mass
increases it causes the large changes in E$_{g}$ (C; LiH; CsH) [30].
Moreover as can be seen from Fig. 3 in last case the empirical dependence ln$%
\partial $E$_{g}$/$\partial $M $\sim $ f(lnE$_{g}$) is very close to the
linear one and in ordinary coordinate system it has a next expression: $%
\sqrt{\partial E_{g}/\partial M\text{ }}$ = E$_{g}^{\sqrt{6.105}}$ ($%
\partial E_{g}/\partial M$ = E$_{g}^{3.0525}$)$.$ By the way it should be
noted that at the large changes E$_{g}$ result in the changes of the force
constants at the isotope effect from the large variation of nuclear mass.
Observable in Fig. 3 rather large scattering data in the rate of change E$%
_{g}$ on the isotope mass in the first step cause the different degree of
the isotope-induced-disorder. The last effect, as is well-known (see
e.g.[4]), due the different magnitude of the scattering potential at the
isotope substitution.

We conclude the discussion of the band-gap shifts with isotope mass by
observing that the effects are exceedingly small in the most cases (see
Table 2) and most likely will not have any consequences any semiconductors
technological applications. However, this exercise is an excellent
demonstration of the advanced state of our quantative theoretical
understanding of the subtle effects of temperature, pressure and isotope
mass on the electronic band structure.

Thanks are due to Dr. M. Segall for a critical reading of the manuscript.

\subsubsection{\textbf{References}}

1. G. Herzberg, Molecular Spectra and Molecular Structure. I. Diatomic
Molecules (Van Nos Reinhold, NY) 1939.

2. V.G. Plekhanov, T.A. Betenekova, V.A. Pustovarov, Sov. Phys. Solid State
18, 1422 (1976); V.G. Plekhanov, Physics-Uspekhi (Moscow) 40, 553 (1997).

3. A.F. Kapustinsky, L.M. Shamovsky and K.S. Baushkina, Physicochim. (USSR)
7, 799 (1937).

4. V.G. Plekhanov, Phys. Rev. B54, 3869 (1996).

5. V.G. Plekhanov, Phys. Solid State (St-Petersburg) 35, 1493 (1993).

6. V.G. Plekhanov, Isotope Effects in Solid State Physics, Academic Press,
NY - London 2001.

7. Y. Onodera and Y. Toyozawa, J. Phys. Soc. Japan 24, 341 (1968).

8. L.F. Lastras-Martinez, T. Ruf, M. Konuma, M. Cardona, Phys. Rev. B61,
12946 (2000).

9. R.J. Elliott, J.A. Krumhansl and P.L. Leath, Rev. Mod. Phys. 46, 465
(1974).

10. A.T. Collins, S.C. Lawson, G. Davies and H. Kanda, Phys. Rev. Lett. 65,
891 (1990).

11. N. Garro, A. Cantarrero, M. Cardona, T. Ruf, Phys. Rev. B54, 4732 (1996).

12. A. Gobel, T. Ruf, Ch. Lin, M. Cardona, Phys. Rev. B56, 210 (1997).

13. V.F. Agekyan, V.M. Asnin, A.M. Kryukov, Fiz. Tverd. Tela 31, 101 (1989)
(in Russian).

14. E.E. Haller, J. Appl. Phys. 77, 2857 (1995).

15. N. Garro, A. Cantarrero, M. Cardona, T. Ruf, Solid State Commun. 98, 27
(1996).

16. F.I. Kreingol'd, Fiz. Tverd. Tela 20, 3138 (1978) (in Russian).

17. F.I. Kreingol'd, ibid, 27, 2839 (1985) (in Russian).

18. F.I. Kreingol'd, K.F. Lider, and L.E. Solov'ev, JETP Lett. 23, 679
(1976) (in Russian).

19. F.I. Kreingol'd, K.F. Lider, and V.F. Sapega, Fiz. Tverd. Tela 19, 3158
(1977) (in Russian).

20. F. I. Kreingol'd and B.S. Kulinkin, ibid 28, 3164 (1986) (in Russian).

21. F.I. Kreingol'd, K.F. Lider, and M.B. Shabaeva, ibid 26, 3940 (1984) (in
Russian).

22. J.M. Zhang, T. Ruf, R. Lauck and M. Cardona, Phys. Rev. B57, 9716 (1998).

23. N. Garro, A. Cantarero, M. Cardona, Phys. Rev. B54, 4732 (1996).

24. P. Etchegoin, J. Weber, M. Cardona, Solid State Commun. 83, 843 (1992).

25. G. Davies, E.C. Lightowlerst, K. Itoh, E.E. Haller, Semicond. Sci.
Technol. 7, 1271 (1992).

26. S. Zollner, M. Cardona and S. Gopalan, Phys. Rev. B45, 3376 (1992).

27. G. Davies, E.C. Lightowlerst, E. E. Haller, Semicond. Sci. Technol. 8,
2201 (1993).

28. C. Parks, A.K. Ramdas, S. Rodriguez, E.E. Haller, Phys. Rev. B49, 14244
(1994).

29. K. Ghanehari, H. Luo, A.L. Ruoff, Solid State, Commun. 95, 385 (1995);
100, 777 (1996).

30. V.G. Plekhanov, Rep. Prog. Phys. 61, 1045 (1998).

\subsubsection{\textbf{Figure Captions.}}

Fig. 1. Mirror reflection spectra of crystals: LiH, curve 1; LiH$_{x}$D$%
_{1-x}$, curve 2; and LiD, curve 3 at 4.2 K. Light source without crystal,
curve 4. Spectral resolution of instrument indicated in the diagram. All
curves have the same energy scale.

Fig. 2. Dependence of the interband transition energy E$_{g}$ in mixed LiH$%
_{x}$D$_{1-x}$ crystals on the hydrogen concentration x. The stright dashed
line is the linear dependence E$_{g}$ = f(x) in the virtual crystal model.
The solid line corresponds to calculations using the polynom of second
degree [2] Points derived from reflection spectra indicated by crosses, and
those from luminescence spectra by triangles.

Fig. 3. The dependence of ln($\partial $E$_{g}$/$\partial $M) $\sim $f(lnE$%
_{g}$); points are experimental data from Table 2 and continuous line -
calculation on the fotmulae (5).

Table 1 .Values of the energy of maxima in exciton reflection spectra of
pure and mixed crystals at 2K, and energies of exciton binding E$_{b}$
band-to-band transitions E$_{g}$ .

\begin{tabular}{|cccccc|}
\hline
\multicolumn{1}{|c|}{Energy, meV} & \multicolumn{1}{c|}{LiH} & 
\multicolumn{1}{c|}{LiH$_{0.82}$D$_{0.18}$} & \multicolumn{1}{c|}{LiH$_{0.40}
$D$_{0.60}$} & \multicolumn{1}{c|}{LiD} & Li$^{6}$H (78K) \\ \hline
\multicolumn{1}{|c|}{E$_{1s}$} & \multicolumn{1}{c|}{4950} & 
\multicolumn{1}{c|}{4967} & \multicolumn{1}{c|}{5003} & \multicolumn{1}{c|}{
5043} & 4939 \\ \hline
\multicolumn{1}{|c|}{E$_{2s}$} & \multicolumn{1}{c|}{4982} & 
\multicolumn{1}{c|}{5001} & \multicolumn{1}{c|}{5039} & \multicolumn{1}{c|}{
5082} & 4970 \\ \hline
\multicolumn{1}{|c|}{E$_{b}$} & \multicolumn{1}{c|}{42} & 
\multicolumn{1}{c|}{45} & \multicolumn{1}{c|}{48} & \multicolumn{1}{c|}{52}
& 41 \\ \hline
\multicolumn{1}{|c|}{E$_{g}$} & \multicolumn{1}{c|}{4992} & 
\multicolumn{1}{c|}{5012} & \multicolumn{1}{c|}{5051} & \multicolumn{1}{c|}{
5095} & 4980 \\ \hline
\end{tabular}

Table 2. Values of the coefficients $\partial $E/$\partial $M (meV) and
energies 0f the band-to-band transitions E$_{g}$(eV) according to indicated
references.

\begin{tabular}{lll}
Substance & $\partial $E$_{g}/\partial $M (meV) & E$_{g}$ (eV) \\ 
$^{13}$C$\rightarrow ^{12}$C & 14.6 [10] & 5.4125 [10] \\ 
$^{7}$LiH$\rightarrow ^{7}$LiD & 103 [2,4] & 4.992$\rightarrow $5.095 [2,4]
\\ 
$^{7}$LiH$\rightarrow ^{6}$LIH & 12 [5] & 4.980 [2,4] \\ 
CsD$\rightarrow $CsH & 60 [29] & 4.440 [29] \\ 
$^{30}$Si$\rightarrow ^{28}$Si & 2 [8] & 3.652 [8] \\ 
$^{68}$ZnO$\rightarrow ^{64}$ZnO & 0.372 [16,20] & 3.400 [16,20] \\ 
Zn$^{18}$O$\rightarrow $Zn$^{16}$O & 3.533 [16,20] & 3.400 [16,20] \\ 
$^{65}$CuCl$\rightarrow ^{63}$CuCl & -0.076 [11,12] & 3.220 [11,12] \\ 
Cu$^{37}$Cl$\rightarrow $Cu$^{35}$Cl & 0.364 [11,12] & 3.220 [11,12] \\ 
Cd$^{34}$S$\rightarrow $Cd$^{32}$S & 0.370 [21] & 2.580 [21] \\ 
$^{110}$CdS$\rightarrow ^{116}$CdS & 0.040$\div $0.068 [22] & 2.580 [22] \\ 
Cu$_{2}^{18}$O$\rightarrow $Cu$_{2}^{16}$O & 1.116 [17-19] & 2.151 [17-19]
\\ 
$^{71}$GaAs$\rightarrow ^{69}$GaAs & 0.39 [15] & 1.53 [15] \\ 
$^{76}$Ge$\rightarrow ^{72}$Ge & 0.225 [13] & 1.53 [13] \\ 
$^{76}\rightarrow ^{73}\rightarrow ^{70}$Ge & 0.37 [24-28] & 0.74 [24-28]%
\end{tabular}

\end{document}